\documentclass[12pt]{article}

\usepackage{amsfonts}

\makeatletter
\def\numberbysection{\@addtoreset{equation}{section}
        \def\theequation{\thesection.\arabic{equation}}}

\def\eqnarray{\stepcounter{equation}\let\@currentlabel=\theequation
\global\@eqnswtrue
\global\@eqcnt\z@\tabskip\@centering\let\\=\@eqncr
$$\halign to \displaywidth\bgroup\@eqnsel\hskip\@centering
  $\displaystyle\tabskip\z@{##}$&\global\@eqcnt\@ne 
  \hfil$\displaystyle{\hbox{}##\hbox{}}$\hfil
  &\global\@eqcnt\tw@ $\displaystyle\tabskip\z@
  {##}$\hfil\tabskip\@centering&\llap{##}\tabskip\z@\cr}
\def\lefteqn#1{\hbox to 2em{$\displaystyle #1$\hss}}
\def\q#1{\raisebox{-0.3ex}{$\displaystyle
  \mathop{q}^{\scriptscriptstyle (#1)}$}{}}

\mathchardef\by="202
\def\mbar#1{\kern 0.1em\overline{\kern -0.1em #1 \kern -0.1em} 
  \kern 0.1em}
\def\invsep{\hskip -2\arraycolsep}
\def\vissep{\hskip -\arraycolsep \vrule \hskip -\arraycolsep}
\makeatother
\numberbysection

\begin{document}

\author{A. V. Razumov, M. V. Saveliev, A. B. Zuevsky \\
{\small \it Institute for High Energy Physics, 142284 Protvino, Moscow
Region, Russia}}
\title{Nonabelian Toda equations\\ associated with classical Lie
groups}
\date{}

\maketitle

\begin{abstract}
The grading operators for all nonequivalent $\mathbb Z$-gradations of
classical Lie algebras are represented in the explicit block matrix
form.
The explicit form of the corresponding nonabelian Toda equations is
given.
\end{abstract}

\section{Introduction}

Toda equations arise in many problems of modern theoretical and
mathematical physics. There is a lot of papers devoted to classical
and
quantum behaviour of abelian Toda equations. From the other hand,
nonabelian Toda equations have not yet received a due attention. From
our
point of view, this is mainly caused by the fact that despite of their
formal exact integrability till recent time there were no nontrivial
examples of nonabelian Toda equations for which one can write the
general
solution in a more or less explicit form. Moreover, even the form of
nonabelian Toda equations was known only for a few partial cases. In
our
recent paper \cite{RSa97} we described some class of nonabelian Toda
equations called there maximally nonabelian. These equations have a
very
simple structure and their general solution can be explicitly written.
Shortly after that we realised that the approach used in \cite{RSa97}
allows to describe the explicit form of all nonabelian Toda equations
associated with classical Lie groups. This is done in the present
paper.

\section{$\mathbb Z$-gradations and Toda equations}

\subsection{Toda equations}

From the point of view of the group-algebraic approach
\cite{LSa92,RSa96}
Toda equations are specified by a choice of a real or complex Lie
group
whose Lie algebra is endowed with a $\mathbb Z$-gradation. Recall that
a
Lie algebra $\mathfrak g$ is said to be $\mathbb Z$-graded, or endowed
with
a $\mathbb Z$-gradation, if there is given a representation of
$\mathfrak
g$ as a direct sum
\[
\mathfrak g = \bigoplus_{m \in \mathbb Z} \mathfrak g_m,
\]
where $[\mathfrak g_m, \mathfrak g_n] \subset \mathfrak g_{m+n}$ for
all
$m, n \in \mathbb Z$.

Let $G$ be a real or complex Lie group, and $\mathfrak g$ be its Lie
algebra. For a given $\mathbb Z$-gradation of $\mathfrak g$ the
subspace
$\mathfrak g_0$ is a subalgebra of $\mathfrak g$. The subspaces
\[
\mathfrak g_{<0} = \bigoplus_{m < 0} {\mathfrak g}_m, \qquad
\mathfrak g_{>0} = \bigoplus_{m > 0} {\mathfrak g}_m
\]
are also subalgebras of $\mathfrak g$. Denote by $G_0$, $G_{<0}$ and
$G_{>0}$ the connected Lie subgroups of $G$ corresponding to the
subalgebras $\mathfrak g_0$, and $\mathfrak g_{<0}$ and $\mathfrak
g_{>0}$
respectively.  

Let $M$ be either the real manifold $\mathbb R^2$ or the complex
manifold
$\mathbb C$. For $M = \mathbb R^2$ we denote the standard coordinates
by
$z^-$ and $z^+$. In the case of $M = \mathbb C$ we use the notation
$z^-$
for the standard complex coordinate and $z^+$ for the complex
conjugate of
$z^-$. Denote the partial derivatives over $z^-$ and $z^+$ by
$\partial_-$
and $\partial_+$ respectively. Consider a Lie group $G$ whose Lie
algebra $\mathfrak g$ is endowed with a $\mathbb Z$-gradation.
Let $l$ be a positive integer, such that the grading subspaces
$\mathfrak
g_m$ for $-l < m < 0$ and $0 < m < l$ are trivial, and $c_-$ and $c_+$
be
some fixed mappings from $M$ to $\mathfrak g_{-l}$ and $\mathfrak
g_{+l}$,
respectively, satisfying the relations
\[
\partial_+ c_- = 0, \qquad \partial_- c_+ = 0.
\]
Restrict ourselves to the case when $G$ is a matrix Lie group. In this
case
the Toda equations are matrix partial differential equations of the
form 
\begin{equation}
\partial_+(\gamma^{-1} \partial_- \gamma) = [c_-, \gamma^{-1} c_+
\gamma], \label{2.2}
\end{equation}
where $\gamma$ is a mapping from $M$ to $G_0$. If the Lie group $G_0$
is
abelian we say that we deal with abelian Toda equations, otherwise we
call them nonabelian Toda equations.

There is a constructive procedure of obtaining the general solution to
Toda
equations \cite{RSa96,RSa96a}. It is based on the use of the Gauss
decomposition related to the $\mathbb Z$-gradation under
consideration.
Here the Gauss decomposition is the representation of an element of
the Lie
group $G$ as a product of elements of the subgroups $G_{<0}$, $G_{>0}$
and
$G_0$ taken in an appropriate order. Another approach is based on the
theory of representations of Lie groups \cite{LSa92,RSa96}. 

\subsection{$\mathbb Z$-gradations of complex semisimple Lie algebras}

Let $q$ be an element of a Lie algebra $\mathfrak g$ such that the
linear
operator \hbox{${\rm ad} \, q$} is semisimple and has integer
eigenvalues.
Defining
\[
\mathfrak g_m = \{x \in \mathfrak g \mid [q, x] = mx\}
\]
we get a $\mathbb Z$-gradation of $\mathfrak g$. This gradation is
said to
be generated by the grading operator $q$. If $\mathfrak g$ is a finite
dimensional complex semisimple Lie algebra, then any $\mathbb
Z$-gradations
of $\mathfrak g$ is generated by a grading operator. Here up to the
action
of the group of the automorphisms of $\mathfrak g$ all $\mathbb
Z$-gradations of $\mathfrak g$ can be obtained with the help of the
following procedure.

Let $\Delta$ be the set of roots of a complex semisimple Lie algebra
$\mathfrak g$ with respect to a Cartan subalgebra $\mathfrak h$, and
$\Pi =
\{\alpha_1, \ldots, \alpha_r\}$ be a base of $\Delta$. Assign to the
vertices of the Dynkin diagram of $\mathfrak g$ nonnegative integer
labels
$q_i$, $i = 1, \ldots, r$, and define
\begin{equation}
q = \sum_{i,j=1}^r h_i (k^{-1})_{ij} q_j \label{2.5}
\end{equation}
where $h_i$ are the corresponding Cartan generators and $k = (k_{ij})$
is
the Cartan matrix of $\mathfrak g$. It is clear that $q$ is a grading
operator of some $\mathbb Z$-gradation of $\mathfrak g$. Here the
subspace
$\mathfrak g_m$ for $m \ne 0$ is the direct sum of the root subspaces
$\mathfrak g^\alpha$ corresponding to the roots $\alpha = \sum_{i=1}^r
n_i
\alpha_i$ with $\sum_{i=1}^r n_i q_i = m$. The subspace $\mathfrak
g_0$,
besides the root subspaces corresponding to the roots $\alpha =
\sum_{i=1}^r n_i \alpha_i$ with $\sum_{i=1}^r n_i q_i = 0$,
includes the Cartan subalgebra $\mathfrak h$.

If all labels $q_i$ are different from zero, then the subgroup
$\mathfrak
g_0$ coincides with the Cartan subalgebra of $\mathfrak h$. In this
case
the subgroup $G_0$ is abelian. In all other cases the subgroup $G_0$
is
nonabelian. The maximally nonabelian Toda equations \cite{RSa97} arise
in
the case when only one of the labels $q_i$ is different from zero. 

\subsection{Conformal invariance}

Let again $G$ be a real or complex Lie group, $\mathfrak g$ be its Lie
algebra, and $M$ be either the real manifold $\mathbb R^2$ or the
complex
manifold $\mathbb C$. Since $M$ is simply connected a connection on
the
trivial principal $G$-bundle $M \times G$ can be identified with a
$\mathfrak g$-valued 1-form $\omega$ on $M$. Here the connection is
flat if
and only if
\begin{equation}
{\rm d} \omega + \omega \wedge \omega = 0. \label{2.16}
\end{equation}
We call this relation the zero curvature condition. It can be shown
that
the Toda equations coincide with the zero curvature condition for the
connection
\begin{equation}
\omega = {\rm d} z^- (\gamma^{-1} \partial_- \gamma + c_-) + {\rm d}
z^+
\gamma^{-1} c_+ \gamma. \label{2.18}
\end{equation}

Let $\xi_\pm$ be some mappings from $M$ to $G_0$, satisfying the
condition
\[
\partial_+ \xi_- = 0, \qquad \partial_- \xi_+ = 0,
\]
and $\gamma$ be a solution of the Toda equations (\ref{2.2}). It is
easy to
get convinced that the mapping 
\begin{equation}
\gamma' = \xi_+^{-1} \gamma \xi_- \label{2.15}
\end{equation}
satisfies the Toda equations (\ref{2.2}) with the mappings $c_\pm$
replaced by the mappings 
\[
c'_\pm = \xi_\pm^{-1} c_\pm \xi_\pm.
\]
In this sense, the Toda equations determined by the mappings $c_\pm$
and $c'_\pm$ which are connected by the above relation, are
equivalent. If the mappings $\xi_\pm$ are such that
\[
\xi_\pm^{-1} c_\pm \xi_\pm = c_\pm,
\]
then transformation (\ref{2.15}) is a symmetry transformation for the
Toda
equations.

Let us show that if the $\mathbb Z$-gradation under consideration is
generated by a grading operator and $c_-$ and $c_+$ are constant
mappings, then the corresponding Toda equations are conformally
invariant.
Let $F: M \to M$ be a conformal transformation. It means that
for the functions $F^- = z^- \circ F$ and $F^+ = z^+ \circ F$ one has
\[
\partial_+ F^- = 0, \qquad \partial_- F^+ = 0.
\]
For the connection $\omega$, given by (\ref{2.18}), we get
\[
F^* \omega = {\rm d} z^- [(\gamma \circ F)^{-1} \partial_- (\gamma
\circ F)
+ \partial_- F^- c_-] + {\rm d} z^+ (\gamma \circ F)^{-1}\partial_+
F^+ c_+
(\gamma \circ F).
\]
If the connection $\omega$ satisfies the zero curvature condition
(\ref{2.16}), then the connection $F^*\omega$ also satisfies this
condition. So if the mapping $\gamma$ satisfies the Toda equations,
then
the mapping $\gamma \circ F$ satisfies the equations
\[
\partial_+[(\gamma \circ F)^{-1} \partial_- (\gamma \circ F)] =
\partial_-
F^- \partial_+ F^+ [c_-, (\gamma \circ F)^{-1} c_+ (\gamma \circ F)].
\]
It is always possible to compensate the factor $\partial_- F^-
\partial_+
F^+$ in the right hand side of the above equation with the help of
transformation (\ref{2.15}). Indeed, defining
\[
\xi_- = \exp \left(- q \, l^{-1} \ln \partial_- F^- \right), \qquad
\xi_+ =
\exp \left( q \, l^{-1} \ln \partial_+ F^+ \right),
\]
one obtains
\[
\xi_-^{-1} c_- \xi_- = (\partial_- F^-)^{-1} c_-, \qquad \xi_+^{-1}
c_+
\xi_+ = (\partial_+ F^+)^{-1} c_+.
\]
Therefore, the mapping
\begin{equation}
\gamma' = \exp \left( - q \, l^{-1} \ln \partial_+ F^+ \right) (\gamma
\circ F) \exp \left( - q \, l^{-1} \ln \partial_- F^- \right).
\label{2.17}
\end{equation}
satisfies the initial Toda equations. Thus, transformation
(\ref{2.17}) is
a symmetry transformation for the Toda equations. Such transformations
define an action of the group of conformal transformations on the
space of
solutions of the Toda equations under consideration. 

\section{Complex general linear group}

We begin the consideration of nonabelian Toda systems associated with
classical Lie groups with the case of the Lie group ${\rm SL}(r+1,
\mathbb
C)$. Actually it is convenient to consider the Lie group ${\rm
GL}(r+1,
\mathbb C)$ whose Lie algebra $\mathfrak{gl}(r+1, \mathbb C)$ is
endowed
with $\mathbb Z$-gradations induced by $\mathbb Z$-gradations of the
Lie
algebra $\mathfrak{sl}(r+1, \mathbb C)$.

The Lie algebra $\mathfrak{sl}(r+1, \mathbb C)$ is of type $A_r$. The
Cartan matrix is
\[
k = \left(\begin{array}{rrrcrrcr}
2 & -1 & 0 & \cdots & 0 & 0 & \invsep & 0 \\
-1 & 2 & -1 & \cdots & 0 & 0 & \invsep & 0 \\
0 & -1 & 2 & \cdots & 0 & 0 & \invsep & 0 \\
\vdots & \vdots & \vdots & \ddots & \vdots & \vdots & \invsep & \vdots
\\
0 & 0 & 0 & \cdots & 2 & -1 & \invsep & 0 \\
0 & 0 & 0 & \cdots & -1 & 2 & \invsep & -1 \\
\cline{7-8}
0 & 0 & 0 & \cdots & 0 & -1 & \vissep & 2
\end{array}\right).
\]
For the inverse matrix one obtains the expression
\[
k^{-1} = \frac{1}{r+1} \left(\begin{array}{ccccccccccc}
r & \invsep & r-1 & \invsep & r-2 & \cdots & 3 & \invsep & 2 & \invsep
& 1
\\
\cline{2-10}
r-1 & \vissep & 2(r-1) & \invsep & 2(r-2) & \cdots & 6 & \invsep & 4 &
\vissep & 2 \\
\cline{5-7}
r-2 & \vissep & 2(r-2) & \vissep & 3(r-2) & \cdots & 9 & \vissep & 6 &
\vissep & 3 \\
\vdots & \vissep & \vdots & \vissep & \vdots & \ddots & \vdots &
\vissep &
\vdots & \vissep & \vdots\\
3 & \vissep & 6 & \vissep & 9 & \cdots & 3(r-2) & \vissep & 2(r-2) &
\vissep & r-2 \\
\cline{5-7}
2 & \vissep & 4 & \invsep & 6 & \cdots & 2(r-2) & \invsep & 2(r-1) &
\vissep & r-1 \\
\cline{2-10}
1 & \invsep & 2 & \invsep & 3 & \cdots & r-2 & \invsep & r-1 & \invsep
& r
\end{array}\right).
\]

Let $d$ be a fixed integer such that $1 \le d \le r$. Consider the
$\mathbb Z$-gradation of $\mathfrak{sl}(r+1, \mathbb C)$ arising when
we
choose the labels of the corresponding Dynkin diagram equal to
zero except the label $q_d$ which is chosen equal to~1. From relation
(\ref{2.5}) it follows that the corresponding grading operator, which
we
denote by $\q{d}$, has the form
\[
\q{d} = \frac{1}{r+1} \left[(r+1-d) \sum_{i=1}^{d-1} i h_i + d
\sum_{i=d}^r (r+1-i) h_i \right].
\]
It is convenient to take as a Cartan subalgebra of $\mathfrak{sl}(r+1,
\mathbb C)$ the subalgebra consisting of diagonal $(r+1) \by (r+1)$
matrices with zero trace. Here the standard choice of the Cartan
generators is 
\[
h_i = e_{i,i} - e_{i+1, i+1},
\]
where the matrices $e_{i, j}$ are defined by
\begin{equation}
(e_{i,j})_{kl} = \delta_{ik} \delta_{jl}. \label{3.34}
\end{equation}
With such a choice of Cartan generators we obtain
\[
\q{d} = \frac{1}{r+1} \left[(r+1-d) \sum_{i=1}^d e_{i,i} - d
\sum_{i=d+1}^{r+1} e_{i,i} \right].
\]
Thus, the grading operator has the following block matrix form:
\[
\q{d} = \frac{1}{r+1} \left( \begin{array}{cc}
k_2 I_{k_1} & 0 \\
0  & -k_1 I_{k_2}
\end{array} \right),
\]
where $k_1 = d$ and $k_2 = r+1-d$, so that $k_1 + k_2 = r+1$.  Here
and henceforth $I_k$ denotes the unit $k \by k$ matrix.

The grading operator corresponding to the general $\mathbb
Z$-gradation of
$\mathfrak{sl}(r+1, \mathbb C)$ is a linear combination of the
operators
$\q{d}$ with nonnegative integer coefficients. The explicit matrix
form of
the grading operators is depicted as follows. A general set of grading
labels $q_i$ can be represented as
\[
(\,\underbrace{0, \ldots, 0}_{k_1 - 1}, m_1, \underbrace{0, \ldots,
0}_{k_2-1}, m_2, 0, \ldots, 0, m_{p-1}, \underbrace{0, \ldots,
0}_{k_{p}-1}\,),
\]
where $k_1, \ldots, k_p$ and $m_1, \ldots, m_{p-1}$ are positive 
integers. It is convenient to consider an arbitrary matrix $x$ of
$\mathfrak{sl}(r+1, \mathbb C)$ as a $p \by p$ block matrix
$(x_{ab})$,
where $x_{ab}$ is a $k_a \by k_b$ matrix. The grading operator
corresponding to the above set of labels has the following block
matrix
form:
\begin{equation}
q = \left( \begin{array}{ccccc}
\rho_1 I_{k_1} & 0 & \cdots & 0 & 0\\
0 & \rho_2 I_{k_2} & \cdots & 0 & 0\\
\vdots & \vdots &\ddots & \vdots & \vdots \\
0 & 0 & \cdots & \rho_{p-1} I_{k_{p-1}} & 0 \\
0 & 0 & \cdots & 0 & \rho_p I_{k_p}
\end{array} \right), \label{3.10}
\end{equation}
where
\[
\rho_a = \frac{1}{r+1} \left( - \sum_{b=1}^{a-1} m_b \sum_{c=1}^b k_c
+
\sum_{b=a}^{p-1} m_b \sum_{c=b+1}^p k_c \right).
\]

We will use grading operator (\ref{3.10}) to define a $\mathbb
Z$-gradation
of the Lie algebra \hbox{$\mathfrak{gl}(r+1, \mathbb C)$}. It is easy
to
describe the arising grading subspaces of $\mathfrak{gl}(r+1, \mathbb
C)$
and the relevant subgroups of ${\rm GL}(r+1, \mathbb C)$. For fixed $a
\ne
b$, the block matrices $x$ having only the block $x_{ab}$ different
from
zero belong to the grading subspace $\mathfrak g_m$ with
\[
m = \sum_{c=a}^{b-1} m_c, \quad a < b, \qquad m = \sum_{c = b}^{a-1}
m_c,
\quad a > b.
\]
The block diagonal matrices form the subalgebra $\mathfrak g_0$.
The subalgebras $\mathfrak g_{<0}$ and $\mathfrak g_{>0}$ are formed
by all
block strictly lower and upper triangular matrices respectively. It is
not
difficult to describe the corresponding subgroups. The subgroup $G_0$
consists of all block diagonal nondegenerate matrices, and the
subgroups
$G_{<0}$ and $G_{>0}$ consist, respectively, of all block lower and
upper
triangular matrices with unit matrices on the diagonal. Note that the
subgroup $G_0$ is isomorphic to the Lie group \hbox{${\rm GL}(k_1,
\mathbb C) \times \cdots \times {\rm GL}(k_p, \mathbb C)$}.

Proceed now to the consideration of the corresponding Toda equations.
Assume that all integers $m_a$ are equal to one. In this case one has
\[
\rho_a = \frac{1}{r+1} \sum_{b=1}^p b k_b - a.
\]
The elements $c_-$ and $c_+$ should belong to the subspaces $\mathfrak
g_{-1}$ and $\mathfrak g_{+1}$ respectively. The general form of such
elements is
\begin{equation}
c_- = \left( \begin{array}{ccccc}
0 & 0 & \cdots & 0 & 0 \\
C_{-1} & 0 & \cdots & 0 & 0 \\
\vdots & \vdots & \ddots & \vdots & \vdots \\
0 & 0 & \cdots & 0 & 0 \\
0 & 0 & \cdots & C_{-(p-1)} & 0
\end{array} \right), \quad 
c_+ = \left( \begin{array}{ccccc}
0 & C_{+1} & \cdots & 0 & 0 \\
0 & 0 & \cdots & 0 & 0 \\
\vdots & \vdots & \ddots & \vdots & \vdots \\
0 & 0 & \cdots & 0 & C_{+(p-1)} \\
0 & 0 & \cdots & 0 & 0
\end{array} \right), \label{3.35}
\end{equation}
where for each $a = 1, \ldots, p-1$ the mapping $C_{-a}$ takes values
in
the space of $k_{a+1} \by k_a$ complex matrices, and the mapping
$C_{+a}$
takes values in the space of $k_a \by k_{a+1}$ complex matrices.
Besides,
these mappings should satisfy the relations
\[
\partial_+ C_{-a} = 0, \qquad \partial_- C_{+a} = 0.
\]
Parametrise the mapping $\gamma$ as
\begin{equation}
\gamma = \left( \begin{array}{ccccc}
\beta_1 & 0 & \cdots & 0 & 0 \\
0 & \beta_2 & \cdots & 0 & 0 \\
\vdots & \vdots & \ddots & \vdots \\
0 & 0 & \cdots & \beta_{p-1} & 0 \\
0 & 0 & \cdots & 0 & \beta_p
\end{array} \right), \label{3.2}
\end{equation}
where the mappings $\beta_a$ take values in the groups ${\rm GL}(m_a,
\mathbb C)$. In this parametrisation Toda equations (\ref{2.2}) take
the form
\begin{eqnarray}
&&\partial_+ (\beta_1^{-1} \partial_- \beta_1) = -\beta_1^{-1} C_{+1}
\beta_{2} C_{-1}, \label{3.3} \\
&&\partial_+ (\beta_a^{-1} \partial_- \beta_a) = -\beta_a^{-1} C_{+a}
\beta_{a+1} C_{-a} + C_{-(a-1)} \beta_{a-1}^{-1} C_{+(a-1)} \beta_a,
\quad
1 < a < p, \hspace{2em} \label{3.4} \\
&&\partial_+ (\beta_p^{-1} \partial_- \beta_p) = C_{-(p-1)}
\beta_{p-1}^{-1} C_{+(p-1)} \beta_p. \label{3.5}
\end{eqnarray}
The consideration of more general $\mathbb Z$-gradations gives nothing
new.
Indeed, the $\mathbb Z$-gradations with all integers $m_a$ equal to 1
exhaust all possible subgroups $G_0$. Furthermore, the mappings
$c_\pm$
corresponding to a general $\mathbb Z$-gradations should take values
in
subalgebras $\mathfrak g_{\pm l}$, where $l$ is less or equal to the
minimal value of the positive integers $m_a$. It is clear that the
blocks $(c_\pm)_{ab}$ are nonzero only if $|a-b|=1$. Therefore, the
general form of the mappings $c_\pm$ is again given by (\ref{3.35}),
where the mappings $C_{\pm a}$ corresponding to the grading indexes
greater than $l$ should be zero mappings.

\section{Complex orthogonal group}

It is convenient for our purposes to define the complex orthogonal
group ${\rm O}(n, \mathbb C)$ as the Lie subgroup of the Lie group
${\rm GL}(n, \mathbb C)$ formed by matrices $a \in {\rm GL}(n, \mathbb
C)$ satisfying the condition 
\begin{equation}
\tilde I_n a^t \tilde I_n = a^{-1}, \label{4.11}
\end{equation} 
where $\tilde I_n$ is the antidiagonal unit $n \by n$ matrix, and
$a^t$ is the transpose of $a$.  The corresponding Lie algebra
$\mathfrak o(n, \mathbb C)$ is the subalgebra of $\mathfrak{gl}(n,
\mathbb C)$ which consists of the matrices $x$ satisfying the
condition
\begin{equation}
\tilde I_n x^t \tilde I_n = -x. \label{4.12}
\end{equation}
For a $k_1 \by k_2$ matrix $a$ we will denote by $a^T$ the matrix
defined by the relation
\[
a^T = \tilde I_{k_2} a^t \tilde I_{k_1}.
\]
Using this notation, we can rewrite conditions (\ref{4.11}) and
(\ref{4.12}) as $a^T = a^{-1}$ and $x^T = -x$.  The Lie algebra
$\mathfrak o(n, \mathbb C)$ is simple.  For $n = 2r+1$ it is of type
$B_r$, while for $n = 2r$ it is of type $D_r$. Discuss these two
cases separately.

Consider the $\mathbb Z$-gradation of $\mathfrak o(2r+1, \mathbb C)$
arising when we choose $q_d = 1$ for some fixed $d$ such that $1 \le
d \le r$, and put all other labels of the Dynkin diagram be equal to
zero. The Cartan matrix for the Lie algebra $\mathfrak o(2r+1, \mathbb
C)$ is given by
\[
k = \left(\begin{array}{rrrcrcrr}
2 & -1 & 0 & \cdots & 0 & \invsep & 0 & 0 \\
-1 & 2 & -1 & \cdots & 0 & \invsep & 0 & 0 \\
0 & -1 & 2 & \cdots & 0 & \invsep & 0 & 0 \\
\vdots & \vdots & \vdots & \ddots & \vdots & \invsep & \vdots & \vdots
\\
0 & 0 & 0 & \cdots & 2 & \invsep & -1 & 0 \\
\cline{7-8}
0 & 0 & 0 & \cdots & -1 & \vissep & 2 & -2 \\
0 & 0 & 0 & \cdots & 0 & \vissep & -1 & 2
\end{array}\right),
\]
and for its inverse one has the expression
\[
k^{-1} = \frac{1}{2} \left(\begin{array}{ccccccccccc}
2 & \invsep & 2 & \invsep & 2 & \cdots & 2 & \invsep & 2 & \vissep & 2
\\
\cline{2-10}
2 & \vissep & 4 & \invsep & 4 & \cdots & 4 & \invsep & 4 & \vissep & 4
\\
\cline{4-10}
2 & \vissep & 4 & \vissep & 6 & \cdots & 6 & \invsep & 6 & \vissep & 6
\\
\vdots & \vissep & \vdots & \vissep & \vdots & \ddots & \vdots &
\invsep &
\vdots & \vissep & \vdots \\
2 & \vissep & 4 & \vissep & 6 & \cdots & 2(r-2) & \invsep & 2(r-2) &
\vissep & 2(r-2) \\
\cline{8-10}
2 & \vissep & 4 & \vissep & 6 & \cdots & 2(r-2) &\vissep & 2(r-1) &
\vissep
& 2(r-1) \\
\cline{1-11}
1 & \invsep & 2 & \invsep & 3 & \cdots & r-2 & \invsep & r-1 & \invsep
& r
\\
\end{array}\right).
\]
Using relation (\ref{2.5}), one gets
\begin{eqnarray*}
&&\q{d} = \sum_{i=1}^{r-1} i h_i + \frac{1}{2} r h_r, \quad d = r, \\
&&\q{d} = \sum_{i=1}^{r-1} i h_i + \frac{1}{2} (r-1) h_r, \quad d =
r-1, \\
&&\q{d} = \sum_{i=1}^d i h_i + d \sum_{i=d+1}^{r-1} h_i + \frac{1}{2}
d
h_r, \quad 1 \le d < r-1.  
\end{eqnarray*}
It is convenient to choose the following Cartan generators of
$\mathfrak
o(2r+1, \mathbb C)$:
\begin{eqnarray*}
&&h_i = e_{i, i} - e_{i+1, i+1} + e_{2r+1-i, 2r+1-i} - e_{2r+2-i,
2r+2-i}, \qquad 1 \le i < r, \\
&&h_r = 2(e_{r,r} - e_{r+2, r+2}),
\end{eqnarray*}
where the matrices $e_{i,j}$ are defined by (\ref{3.34}). Using these
expressions one obtains
\[
\q{d} = \sum_{i=1}^d e_{i, i} - \sum_{i=1}^d e_{2r+2-i, 2r+2-i}. 
\]
Denoting $k_1 = d$ and $k_2 = 2(r-d) + 1$, we write $q$ in block
matrix form,
\begin{equation}
\q{d} = \left( \begin{array}{ccc}
I_{k_1} & 0 & 0 \\
0 & 0 & 0 \\
0 & 0 & -I_{k_1}
\end{array} \right), \label{4.3}
\end{equation}
where zero on the diagonal stands for the $k_2 \by k_2$ block of
zeros.

The Cartan matrix for the Lie algebra $\mathfrak o(2r, \mathbb C)$ has
the
form
\[
k = \left(\begin{array}{rrrccrrr}
2 & -1 & 0 & \cdots & \invsep & 0 & 0 & 0 \\
-1 & 2 & -1 & \cdots & \invsep & 0 & 0 & 0 \\
0 & -1 & 2 & \cdots & \invsep & 0 & 0 & 0 \\
\vdots & \vdots & \vdots &\ddots & \invsep & \vdots & \vdots & \vdots
\\
\cline{5-8}
0 & 0 & 0 & \cdots & \vissep & 2 & -1 & -1 \\
0 & 0 & 0 & \cdots & \vissep & -1 & 2 & 0 \\
0 & 0 & 0 & \cdots & \vissep & -1 & 0 & 2
\end{array}\right),
\]
and its inverse is
\[
k^{-1} = \frac{1}{4} \left(\begin{array}{cccccccccccc}
4 & \invsep & 4 & \invsep & 4 & \cdots & \invsep & 4 & \vissep & 2 &
\vissep & 2 \\
\cline{2-9}
4 & \vissep & 8 & \invsep & 8 & \cdots & \invsep & 8 & \vissep & 4 &
\vissep & 4 \\
\cline{4-9}
4 & \vissep & 8 & \vissep & 12 & \cdots & \invsep & 12 & \vissep & 6 &
\vissep & 6 \\
\vdots & \vissep & \vdots & \vissep & \vdots & \ddots & \invsep &
\vdots &
\vissep & \vdots & \vissep & \vdots \\
\cline{7-9}
4 & \vissep & 8 & \vissep & 12 & \cdots & \vissep & 4(r-2) & \vissep &
2(r-2) & \vissep & 2(r-2) \\
\cline{1-12}
2 & \invsep & 4 & \invsep & 6 & \cdots & \invsep & 2(r-2) &\vissep & r
&
\vissep & r-2 \\
\cline{1-12}
2 & \invsep & 4 & \invsep & 6 & \cdots & \invsep & 2(r-2) & \vissep &
r-2 &
\vissep & r \\
\end{array}\right).
\]
In this case one obtains
\begin{eqnarray*}
&&\q{d} = \frac{1}{2} \sum_{i=1}^{r-2} i h_i + \frac{1}{4} (r-2)
h_{r-1} +
\frac{1}{4} r h_r, \quad d = r, \\
&&\q{d} = \frac{1}{2} \sum_{i=1}^{r-2} i h_i + \frac{1}{4} r h_{r-1} +
\frac{1}{4} (r-2) h_r, \quad d = r-1, \\
&&\q{d} = \sum_{i=1}^d i h_i + d \sum_{i = d+1}^{r-2} h_i +
\frac{1}{2} d
(h_{r-1} + h_r), \quad 1 \le d < r-1. 
\end{eqnarray*}
Choose as the Cartan generators of $\mathfrak o(2r, \mathbb C)$ the
elements 
\begin{eqnarray*}
&&h_i = e_{i,i} - e_{i+1, i+1} + e_{2r-i, 2r-i} - e_{2r+1-i, 2r+1-i},
\quad 1 \le i < r, \\
&&h_r = e_{r-1, r-1} + e_{r,r} - e_{r+1, r+1} - e_{r+2, r+2}.
\end{eqnarray*}
Then it is easy to see that 
\begin{eqnarray*}
&&\q{d} = \frac{1}{2} \sum_{i=1}^r e_{i, i} - \frac{1}{2} \sum_{i=1}^r
e_{2r+1-i, 2r+1-i}, \quad d = r, \\
&&\q{d} = \frac{1}{2} \sum_{i=1}^{r-1} e_{i, i} - \frac{1}{2} e_{r, r}
+
\frac{1}{2} e_{r+1, r+1} - \frac{1}{2} \sum_{i=1}^{r-1} e_{2r+1-i,
2r+1-i}, \quad d = r-1, \\
&&\q{d} = \sum_{i=1}^d e_{i, i} - \sum_{i=1}^d e_{2r+1-i, 2r+1-i},
\quad
1 \le d < r-1. 
\end{eqnarray*}
Note that the grading operators corresponding to the cases $d = r$
and $d = r-1$ are connected by the automorphism $\sigma$ of $\mathfrak
o(2r, \mathbb C)$ defined by the relation $\sigma(x) = axa^{-1}$,
where
\[
a = \left( \begin{array}{cccccccc}
1 & \cdots & 0 & 0 & 0 & 0 & \cdots & 0 \\
\vdots & \ddots & \vdots & \vdots & \vdots & \vdots & \ddots & \vdots
\\
0 & \cdots & 1 & 0 & 0 & 0 & \cdots & 0 \\
0 & \cdots & 0 & 0 & 1 & 0 & \cdots & 0 \\
0 & \cdots & 0 & 1 & 0 & 0 & \cdots & 0 \\
0 & \cdots & 0 & 0 & 0 & 1 & \cdots & 0 \\
\vdots & \ddots & \vdots & \vdots & \vdots & \vdots & \ddots & \vdots
\\
0 & \cdots & 0 & 0 & 0 & 0 & \cdots & 1
\end{array} \right).
\]
There is the corresponding automorphism of the Lie group ${\rm O}(2r,
\mathbb C)$, which is defined by the same formula. Thus, the cases
$d=r$
and $d=r-1$ leads actually to the same $\mathbb Z$-gradation.

For the case $d=r$ the grading operator has the following block form
\begin{equation}
\q{r} = \frac{1}{2} \left( \begin{array}{cc}
I_k & 0 \\
0 & -I_k
\end{array} \right), \label{4.4}
\end{equation}
where we denoted $k = r$. In the case $1 \le d < r-2$ denoting $k_1 =
d$ and $k_2 = 2(r-d)$ one sees that the grading operator $q$ has form
(\ref{4.3}). 

The grading operator of a general $\mathbb Z$-gradation of the Lie
algebra
$\mathfrak o(n, \mathbb C)$ is again a linear combination of the
grading
operators $\q{d}$ with non-negative integer coefficients. Using the
explicit form of the operators $\q{d}$ and taking into account the
existence of the automorphism of $\mathfrak o(2r, \mathbb C)$
described
above, we come to the following explicit description of the $\mathbb
Z$-gradations of $\mathfrak o(n, \mathbb C)$. 

A $\mathbb Z$-gradation of $\mathfrak o(n, \mathbb C)$ is determined
first
by a fixation of block matrix representation of the elements of
$\mathfrak
o(n, \mathbb C)$. Here any element $x$ is seen as a $p \by p$ block
matrix
$(x_{ab})$, where $p \le n$ and $x_{ab}$ is a $k_a \by k_b$ matrix.
Now the
positive integers $k_a$ are not arbitrary. They are restricted by the
relation
\[
k_a = k_{p-a+1}.
\]
To get a concrete $\mathbb Z$-gradation, one also have to fix a set of
positive integers $m_a$, $a=1, \ldots, p-1$, subjected to the
constraint
\[
m_a = m_{p-a}.
\]
The corresponding grading operator has the form (\ref{3.10}) with
\[
\rho_a = \frac{1}{2} \left( - \sum_{b=1}^{a-1} m_b + \sum_{b=a}^{p-1}
m_b
\right).
\]
The structure of the subalgebras $\mathfrak g_0$, $\mathfrak g_{<0}$,
$\mathfrak g_{>0}$ and the corresponding subgroups is the same as in
the
case of general linear group with the exception that we should use
only
those block matrices which belong to ${\rm SO}(n, \mathbb C)$. It is
clear
that the subgroup $G_0$ for an odd $p = 2s+1$ is isomorphic to the Lie
group ${\rm GL}(k_1, \mathbb C) \times \cdots \times {\rm GL}(k_s,
\mathbb
C) \times {\rm SO}(k_{s+1}, \mathbb C)$ while for an even $p = 2s$ it
is
isomorphic to the Lie group ${\rm GL}(k_1, \mathbb C) \times \cdots
\times
{\rm GL}(k_s, \mathbb C)$. Note that the latter is possible only if
$n$ is
even.

Consider now the corresponding Toda equations. As for the case of the
general linear group it suffices to consider only the $\mathbb
Z$-gradations for which all integers $m_a$ are equal to~1. In this
case one
has
\[
\rho_a = \frac{p+1}{2} - a.
\]
The general form of the mappings $c_\pm$ is given by (\ref{3.35})
where
\begin{equation}
C_{\pm a}^T = - C_{\pm(p-a)}^{}. \label{4.5}
\end{equation}
We will use the parametrisation of the mapping $\gamma$ given by
(\ref{3.2}), where
\begin{equation}
\beta_a^T = \beta_{p-a+1}^{-1}. \label{4.6}
\end{equation}
So for the case $p = 2s+1$ we have $s+1$ independent mappings
$\beta_a$ and
in the case $p = 2s$ there are $s$ independent mappings. 

The Toda equations has form (\ref{3.3})--(\ref{3.5}), where the
mappings
$C_{\pm a}$ and $\beta_a$ satisfy relations (\ref{4.5}) and
(\ref{4.6}). In
the case $p = 2s + 1$ for the independent mappings $\beta_1, \ldots,
\beta_{s+1}$ one can write
\begin{eqnarray*}
&&\partial_+ (\beta_1^{-1} \partial_- \beta_1^{}) = -\beta_1^{-1}
C_{+1}^{}
\beta_2^{} C_{-1}^{}, \\
&&\partial_+ (\beta_a^{-1} \partial_- \beta_a^{}) = -\beta_a^{-1}
C_{+a}^{}
\beta_{a+1}^{} C_{-a}^{} + C_{-(a-1)}^{} \beta_{a-1}^{-1}
C_{+(a-1)}^{}
\beta_a^{}, \quad 1 < a \le s, \\
&&\partial_+ (\beta_{s+1}^{-1} \partial_- \beta_{s+1}^{}) = -
\beta_{s+1}^T
C_{+s}^T \beta_s^{-1T} C_{-s}^T + C_{-s}^{} \beta_s^{-1} C_{+s}^{}
\beta_{s+1}^{}.
\end{eqnarray*}
Note that in this case $\beta_{s+1}^T = \beta_{s+1}^{-1}$. In the case
$p =
2s$ the independent equations are
\begin{eqnarray}
&&\partial_+ (\beta_1^{-1} \partial_- \beta_1^{}) = -\beta_1^{-1}
C_{+1}^{}
\beta_2^{} C_{-1}^{}, \label{4.7} \\
&&\partial_+ (\beta_a^{-1} \partial_- \beta_a^{}) = -\beta_a^{-1}
C_{+a}^{}
\beta_{a+1}^{} C_{-a}^{} + C_{-(a-1)}^{} \beta_{a-1}^{-1}
C_{+(a-1)}^{}
\beta_a^{}, \quad 1 < a < s, \hspace{2em} \label{4.8} \\
&&\partial_+ (\beta_{s}^{-1} \partial_- \beta_s^{}) = - \beta_{s}^{-1}
C_{+s}^{} \beta_s^{-1T} C_{-s}^{} + C_{-(s-1)}^{} \beta_{s-1}^{-1}
C_{+(s-1)}^{} \beta_s^{}, \label{4.9}
\end{eqnarray}
where $C_{-s}^T = -C_{-s}^{}$ and $C_{+s}^T = -C_{+s}^{}$.

\section{Complex symplectic group}

We define the complex symplectic group ${\rm Sp}(2r, \mathbb C)$ as
the
Lie subgroup of the Lie group ${\rm GL}(2r, \mathbb C)$ which consists
of the matrices $a \in {\rm GL}(2r, \mathbb C)$ satisfying the
condition
\[
\tilde J_r a^t \tilde J_r  = - a^{-1},
\]
where $\tilde J_r$ is the matrix given by
\[
\tilde J_r = \left( \begin{array}{cc}
0 & \tilde I_r \\
-\tilde I_r & 0
\end{array} \right).
\]
The corresponding Lie algebra $\mathfrak{sp}(r, \mathbb C)$ is defined
as
the subalgebra of the Lie algebra $\mathfrak{sl}(2r, \mathbb C)$
formed
by the matrices $x$ which satisfy the condition
\[
\tilde J_r x^t \tilde J_r = x.       
\]
The Lie algebra $\mathfrak{sp}(r, \mathbb C)$ is simple, and it is of
type $C_r$. Therefore, the Cartan matrix of $\mathfrak{sp}(r, \mathbb
C)$
is the transpose of the Cartan matrix of $\mathfrak o(2r, \mathbb C)$,
and the same is true for the inverse of the Cartan matrix of
$\mathfrak{sp}(r,\mathbb C)$. Thus, the explicit form of the Cartan
matrix
is
\[
k = \left(\begin{array}{rrrcrcrr}
2 & -1 & 0 & \cdots & 0 & \invsep & 0 & 0 \\
-1 & 2 & -1 & \cdots & 0 & \invsep & 0 & 0 \\
0 & -1 & 2 & \cdots & 0 & \invsep & 0 & 0 \\
\vdots & \vdots & \vdots & \ddots & \vdots & \invsep & \vdots & \vdots
\\
0 & 0 & 0 & \cdots & 2 & \invsep & -1 & 0 \\
\cline{7-8}
0 & 0 & 0 & \cdots & -1 & \vissep & 2 & -1 \\
0 & 0 & 0 & \cdots & 0 & \vissep & -2 & 2
\end{array}\right),
\]
and for its inverse one has
\[
k^{-1} = \frac{1}{2} \left(\begin{array}{ccccccccccc}
2 & \invsep & 2 & \invsep & 2 & \cdots & 2 & \invsep & 2 & \vissep & 1
\\
\cline{2-10}
2 & \vissep & 4 & \invsep & 4 & \cdots & 4 & \invsep & 4 & \vissep & 2
\\
\cline{4-10}
2 & \vissep & 4 & \vissep & 6 & \cdots & 6 & \invsep & 6 & \vissep & 3
\\
\vdots & \vissep & \vdots & \vissep & \vdots & \ddots & \vdots &
\invsep &
\vdots & \vissep & \vdots \\
2 & \vissep & 4 & \vissep & 6 & \cdots & 2(r-2) & \invsep & 2(r-2) &
\vissep & r-2 \\
\cline{8-10}
2 & \vissep & 4 & \vissep & 6 & \cdots & 2(r-2) &\vissep & 2(r-1) &
\vissep
& r-1 \\
\cline{1-10}
2 & \invsep & 4 & \invsep & 6 & \cdots & 2(r-2) & \invsep & 2(r-1) &
\vissep & r \\
\end{array}\right).
\]
For any fixed integer $d$ such that $1 \le d \le r$, consider the
$\mathbb
Z$-gradation of $\mathfrak{sp}(r, \mathbb C)$ arising when we choose
all
the labels of the corresponding Dynkin diagram equal to zero, except
the
label $q_d$, which we choose be equal to~$1$.

Using relation (\ref{2.5}), we obtain the following expressions for
the
grading operator,
\[
\q{d} = \frac{1}{2} \sum_{i=1}^r i h_i, \quad d = r, \qquad
\q{d} = \sum_{i=1}^d i h_i + d \sum_{i=d+1}^r h_i, \quad 1 \le d < r.
\]
Using the following choice of the Cartan generators,
\begin{eqnarray*}
&&h_i = e_{i,i} - e_{i+1, i+1} + e_{2r-i, 2r-i} - e_{2r+1-i, 2r+1-i},
\qquad 1 \le i < d, \\
&&h_r = e_{r, r} - e_{r+1, r+1},
\end{eqnarray*}
one sees that the grading operator for the case $d = r$ has form
(\ref{4.4}) with $k = r$, and for the case $1 \le d < r$ it has form
(\ref{4.3}) with $k_1 = d$ and $k_2 = 2(r-d)$. So we have the same
grading
operators and, therefore, the same structure of grading subspaces as
we had
in the case of the Lie algebra $\mathfrak{so}(2r, \mathbb C)$. 
In the case of odd $p = 2s+1$ the subgroup $G_0$ is isomorphic to the
Lie
group ${\rm GL}(k_1, \mathbb C) \times \cdots \times {\rm GL}(k_s,
\mathbb
C) \times {\rm Sp}(k_{s+1}, \mathbb C)$. Note that here $k_{s+1}$ is
even.
In the case $p = 2s$ the subgroup $G_0$ is isomorphic to ${\rm
GL}(k_1,
\mathbb C) \times \cdots \times {\rm GL}(k_s, \mathbb C)$.

Without any loss of generality we assume that all integers $m_a$
characterising the $\mathbb Z$-gradation are equal to 1. The general
form
of the mappings $c_\pm$ is given by (\ref{3.35}), where in the case $p
=
2s+1$ one has
\begin{eqnarray*}
&&C_{-a}^T = -C_{-(p-a)}, \qquad C_{+a}^T = -C_{+(p-a)},\qquad a \ne
s, \\
&&\tilde I_{k_s} C_{-s}^t \tilde J_{k_{s+1}/2} = -C_{-(s+1)}, \qquad
\tilde
J_{k_{s+1}/2} C_{+s}^t \tilde I_{k_s} = C_{+(s+1)}.
\end{eqnarray*}
In the case $p = 2s$ the mappings $C_{\pm a}$ should satisfy the
relations
\begin{eqnarray*}
&&C_{-a}^T = -C_{-(p-a)}, \qquad C_{+a}^T = -C_{+(p-a)},\qquad a \ne
s, \\
&&C_{-s}^T = C_{-s}, \qquad C_{+s}^T = C_{+s}.
\end{eqnarray*}
To write the Toda equations in an explicit form we use again the
parametrisation (\ref{3.2}), where in the case $p = 2s+1$ one has
\begin{eqnarray*}
&&\beta_a^T = \beta_{p-a+1}^{-1}, \qquad a \ne s+1, \\
&&\tilde J^{}_{k_{s+1}/2} \beta_{s+1}^t \tilde J^{}_{k_{s+1}/2} = -
\beta_{s+1}^{-1},
\end{eqnarray*}
whereas in the case $p = 2s$
\[
\beta_a^T = \beta_{p-a+1}^{-1}
\]
for any $a = 1, \ldots, 2s$. The independent Toda equations in the
case
$p=2s+1$ are
\begin{eqnarray*}
&&\partial_+ (\beta_1^{-1} \partial_- \beta_1) = -\beta_1^{-1} C_{+1}
\beta_{2} C_{-1}, \\
&&\partial_+ (\beta_a^{-1} \partial_- \beta_a) = -\beta_a^{-1} C_{+a}
\beta_{a+1} C_{-a} + C_{-(a-1)} \beta_{a-1}^{-1} C_{+(a-1)} \beta_a,
\quad
1 < a \le s, \\
&&\partial_+ (\beta_{s+1}^{-1} \partial_- \beta_{s+1}) =
\beta_{s+1}^{-1}
\tilde J_{k_{s+1}/2} C_{+s}^t \beta_s^{-1t} C_{-s}^t \tilde
J_{k_{s+1}/2} +
C_{-s} \beta_s^{-1} C_{+s} \beta_{s+1}.
\end{eqnarray*}
In the case $p = 2s$ one has equations (\ref{4.7})--(\ref{4.9}), where
$C_{-s}^T = C_{-s}$ and $C_{+s}^T = C_{+s}$.

\section{Concluding remarks}

To construct the general solution for the equations described in the
present paper one can apply the method based on the Gauss
decomposition. For some partial cases this is done in our paper
\cite{RSa97}. The method based on the representation theory was
applied to this problem by A.~N.~Leznov \cite{Lez98a,Lez98b}. One can
also use the methods considered by A.~N.~Leznov and E.~A.~Yusbashjan
\cite{LYu95} and by P.~Etingof, I.~Gelfand and V.~Retakh
\cite{EGR97,EGR97a} which lead to some very simple forms of the
solution but, unfortunately, cannot be applied in general situation.

It is worth to note that all nonabelian Toda equations associated with
the Lie groups ${\rm SO}(n, \mathbb C)$ and ${\rm Sp}(n = 2m, \mathbb
C)$ can be obtained by reduction of appropriate equations associated
with the Lie group ${\rm GL}(n, \mathbb C)$. Actually this fact can be
proved without using concrete matrix realisation of the Lie groups and
Lie algebras under consideration.\footnote{We are thankful to
A.~N.~Leznov for the discussion of this point.}

The results obtained above can be generalised to the case of higher
grading Toda equations \cite{GSa95,Lez97} and multidimensional
Toda-type equations \cite{RSa96a}.

From the point of view of physical applications it is interesting to
investigate possible reductions to real Lie groups. Some results in
this direction valid for $\mathbb Z$-gradations generated by the
Cartan generator of some ${\rm SL}(2, \mathbb C)$-subgroup of $G$ are
obtained by J.~M.~Evans and J.~O.~Madsen \cite{EMa98}.

We believe that nonabelian Toda equations are quite relevant for a
number of problems of theoretical and mathematical physics, and in a
near future their role for the description of nonlinear phenomena in
many areas will be not less than that of the abelian Toda equations.

\section*{Acknowledgements}

It is a pleasure to thank J.-L.~Gervais and Yu.~I.~Manin for many
fruitful discussions. The research program of the authors was
supported in part by the Russian Foundation for Basic Research under
grant \#~98--01--00015 and by INTAS grant\#~96--690.

\end{document}